\documentclass[a4paper]{article}
\usepackage{float}

\usepackage{titlesec}

\usepackage{tcolorbox}
\usepackage{graphicx}
\usepackage{amsmath, amssymb}
\usepackage{multirow}
\usepackage{booktabs}
\usepackage{hyperref}
\usepackage{caption}
\usepackage{subcaption}
\usepackage{float}
\usepackage{cleveref}
\usepackage{tcolorbox}
\usepackage{makecell}
\usepackage{hhline}
\usepackage{adjustbox}
\usepackage{geometry}
\usepackage{url}
\usepackage{enumitem}
\usepackage{xcolor}
\usepackage{amsthm}

\title{Harnessing Fast Fourier Transform for Rapid Community Travel Distance and Step Estimation in Children with Duchenne Muscular Dystrophy}

\author{
    Erik K. Henricson\thanks{\small Corresponding author}\\
    \texttt{ehenricson@ucdavis.edu} \\
    \and
    Albara Ah Ramli \\
    \texttt{arramli@ucdavis.edu}
}

\date{
Department of Physical Medicine and Rehabilitation, University of California, Davis
}

\begin{document}

\maketitle

\begin{abstract}
Accurate estimation of gait characteristics, including step length, step velocity, and travel distance, is critical for assessing mobility in toddlers, children and teens with Duchenne muscular dystrophy (DMD) and typically developing (TD) peers. This study introduces a novel method leveraging Fast Fourier Transform (FFT)-derived step frequency from a single waist-worn consumer-grade accelerometer to predict gait parameters efficiently. The proposed FFT-based step frequency detection approach, combined with regression-derived stride length estimation, enables precise measurement of temporospatial gait features across various walking and running speeds. Our model, developed from a diverse cohort of children aged 3-16, demonstrated high accuracy in step length estimation ($R^2=0.92$, $RMSE = 0.06$) using only step frequency and height as inputs. Comparative analysis with ground-truth observations and AI-driven Walk4Me models validated the FFT-based method, showing strong agreement across step count, step frequency, step length, step velocity, and travel distance metrics. The results highlight the feasibility of using widely available mobile devices for gait assessment in real-world settings, offering a scalable solution for monitoring disease progression and mobility changes in individuals with DMD. Future work will focus on refining model performance and expanding applicability to additional movement disorders.
\end{abstract}




\section{Introduction}

The number of steps taken over time and the length of those steps are fundamental elements of human gait, which together describe both velocity and the distance traveled during locomotion. The relationships between these variables have been studied extensively and form the foundation of modern gait analysis, both for typically developing individuals and those with Duchenne muscular dystrophy (DMD). A detailed review of early and contemporary work on this topic is beyond the scope of this paper, but we refer the reader to studies by G.A. Dean, Inman, Ralston, Todd for typical gait, and Sutherland and D’Angelo for DMD-specific characteristics~\cite{dean1965analysis,inman1981human,sutherland1981pathomechanics,d2009gait}.

Complex mathematical models exist that describe the contributions of each joint and muscle group to gait, as well as how anatomical and functional disturbances can lead to clinically significant mobility limitations, which affect daily life in a variety of patient populations. At its most basic level, however, the velocity and length of each step influences how far and how quickly we move throughout our daily activities.

Recently, the European Medicines Agency (EMA) has endorsed the 95th percentile of maximal step velocity over weeks-long periods of assessment as a clinically meaningful outcome measure of community ambulation. This measure is particularly relevant for clinical trials involving people with DMD~\cite{committee2023qualification}. The ankle-worn device approved for this measurement is elegant, accurate, and well-tolerated, even by young children~\cite{servais2021first, servais2024evidentiary, lilien2019home, poleur2021normative, servais2022stride, rabbia2024stride}. However, it remains a custom, high-cost, laboratory-grade device produced in small quantities, making it difficult and expensive to use in large-scale public health or registry-style studies involving people with DMD.

To address these challenges, we developed methods and software that leverage widely available mobile phones and consumer-level devices to accurately measure steps, estimate step length, walking speed, and travel distance in the everyday lives of individuals with DMD, other mobility disorders, and typically developing individuals. 

In our previous work, we described how we created a dashboard of clinical features extracted from iPhone accelerometer data collected during ambulation using our Walk4Me software suite~\cite{ramli2024gait_1,ramli2024gait_2,ramli2023walk4me}. Our tools extract clinically meaningful temporospatial features of gait to support both classical machine learning and deep learning models. These models are capable of differentiating between gait patterns of children with Duchenne Muscular Dystrophy (DMD) and typically developing peers.

The tools rely on computationally intensive, individually calibrated regression-based models and AI-driven step detection algorithms. The models relate acceleration forces to steps of a given length, providing highly accurate estimates of step counts, frequencies, lengths, velocities, and distances traveled during short-duration clinical tests such as the 10- and 25-meter timed walk/run tests, the 100-meter run, and the 6-minute walk test. While the tools can also be applied to longer bouts of activity, they require significant processing time for such tasks.

To improve the speed and accuracy of feature estimation during longer bouts of community-based travel, we developed a tool that requires only the participant’s height and step frequency as inputs. Inspired by the step length estimation models developed by Dean and Todd, we created a novel step length prediction tool specifically for DMD children and typical controls. This tool is based on multi-speed step length calibration data from our original cohort. Additionally, we used rapid, FFT-based frequency analysis on time-windowed raw accelerometer data to determine step frequency.

In this paper, we present our equations and techniques, evaluating the accuracy and precision of our estimates by comparing them to both ground-truth observations and estimates generated using our original, individually calibrated methods. Specifically, we introduce a novel FFT-based approach to estimate step frequency, step length, step velocity, and total travel distance using a single waist-worn sensor. We validate our method against ground-truth observations and a machine-learning-based approach (Walk4Me), demonstrating strong agreement. Furthermore, we provide a computationally efficient alternative for large-scale, real-world gait monitoring in children with DMD.

\section{Experiment-I: Developing an equation to predict stride length using step frequency and height in children with DMD and typical controls}
\label{Experiment-I}

\subsection{Materials and~Methods}

\subsubsection{Assessment of participants}
We conducted short-distance walking events (very slow, slow, self-selected comforable and fast walks) from ambulatory DMD children and teens typical similar-aged controls. Assessments were conducted and recorded by a trained clinical evaluator on a marked, straight-line course using previously-described methods~\cite{mcdonald20136, escolar2001clinical, mayhew2007reliable}. Participants with DMD walked a 10-meter distance, while typical controls walked a 25-meter course.  Participants completed longer-distance, mixed speed walking and running events including a 100-meter walk/run, 6-minute walk test, and a self-selected pace free walk.  Assessments were recorded on video for later ground-truth validation of step counts and distances. We recorded triaxial acceleration signals using an iPhone running our Walk4Me application, placed at the waist near the lumbosacral junction as previously described~\cite{ramli2024gait_1,ramli2024gait_2,ramli2023walk4me}.

\subsubsection{Predictive modeling of step length in DMD patients and typical controls}

Using the short-distance walking events, we used mixed model regression techniques accounting for multiple measures within participant to explore the relationship between observed step length, step cadence and standing height in participants with DMD and control participants.  We calculated step length as the distance divided by the number of observed steps and cadence as steps per second during each effort. We calculated standing height in meters, and included an indicator variable of (1) for individuals with DMD and (0) for controls. Using these inputs, we explored multiple transformations of the raw data to satisfy assumptions of normality in ordinary least squares regression. We then constructed multiple exploratory models to estimate step length based on independent variables of step cadence, height and DMD status and their combined interactions. We evaluated model performance based on presence of statistically- significant contributing variables and interactions, model $R^2$ values, and minimization of error. 

\subsection{Results}

\subsubsection{Participants}
Participant characteristics are shown in Table 1. We conducted 261 short-distance walking events in 19 ambulatory DMD patients and 25 typical controls over 67 visits.


\begin{table}[htbp]
\centering
\renewcommand{\arraystretch}{1.3}
\caption{Characteristics of the children included in the study.}
\label{participant_table}
\setlength{\tabcolsep}{3pt} 
\begin{adjustbox}{max width=\textwidth}
\begin{tabular}{lccccccccc}
\toprule
\textbf{Measure} & \multicolumn{4}{c}{\textbf{DMD}} & \multicolumn{4}{c}{\textbf{Control}} & \textbf{p-value} \\
\cmidrule(lr){2-5} \cmidrule(lr){6-9}
 & \textbf{N} & \textbf{Mean} & \textbf{SD} & \textbf{Range} & \textbf{N} & \textbf{Mean} & \textbf{SD} & \textbf{Range} & \\
\midrule
Age (years) & 19 & 9.79 & 3.71 & (3.0--16.0) & 25 & 7.6 & 3.33 & (2.0--15.0) & 0.0273 \\ 
Height (cm) & 19 & 127.6 & 14.86 & (101.6--153.3) & 25 & 131.36 & 19.8 & (105.5--175.0) & 0.6181 \\ 
Weight (kg) & 19 & 37.19 & 14.56 & (17.2--67.7) & 25 & 33.26 & 18.01 & (18.5--101.0) & 0.4070 \\ 
NSAA Score & 17 & 20.71 & 7.41 & (8.0--32.0) & 25 & 32.96 & 2.11 & (26.0--34.0) & $<$0.0001 \\ 
10MRW Time (s) & 18 & 6.44 & 2.29 & (3.66--10.6) & 24 & 3.52 & 0.83 & (2.16--6.05) & $<$0.0001 \\ 
6MWT Distance (m) & 18 & 370.22 & 84.11 & (227.0--519.0) & 24 & 541.69 & 123.01 & (233.0--825.5) & $<$0.0001 \\ 
\bottomrule
\end{tabular}
\end{adjustbox}
\end{table}

\subsubsection{Equation for step length prediction}

The best-performing model for step length included coefficients for square root-transformed cadence, inverse square root-transformed height and the DMD status indicator variable, and interaction terms for transformed cadence and height and DMD status. The adjusted $R^2$ for the resulting model was 0.92 (p$<$0.0001), with a root mean squared error of 0.06 indicating good model performance. 

\begin{figure}[htbp] 
\centering
\includegraphics[scale=0.63]{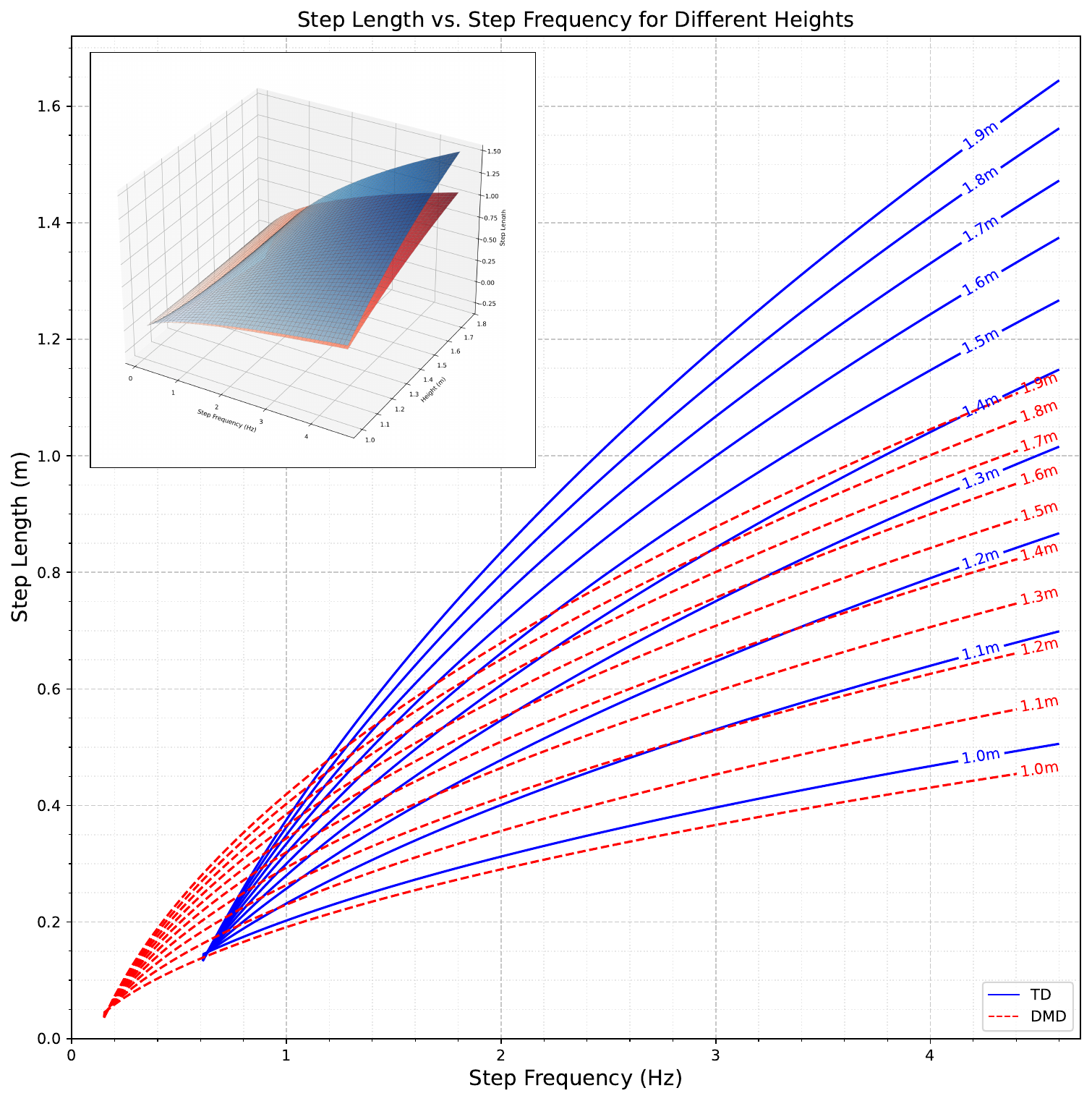}
\caption{Surface plot illustrating the relationship between step frequency, standing height, and step length in children and adolescents with Duchenne muscular dystrophy (DMD) (red) and typically developing controls (blue). The model is based on Equation~\eqref{EQ3_FORMAT}. The curves illustrates the relationship between Step Length and Step Frequency at various heights, as modeled by Equation~\eqref{EQ3_FORMAT}, for TD individuals and those with DMD. The plotted curves represent the regression model, with the blue line depicting TD participants and the red dotted line representing DMD participants.}
\label{FIG_MIX}  
\end{figure}

The model predicting step length ($SL$) takes a reduced form where sf is step frequency in steps per second and h is standing height in meters and $DMD(1)$ serves as the indicator variable (1=DMD, 0=Control) that is applied if the individual has a DMD diagnosis. A surface plot of Equation~\eqref{EQ3_FORMAT} is shown in Figure~\ref{FIG_MIX}, below.

\scriptsize
\begin{equation} \label{EQ3_FORMAT}
\begin{split}
SL(sf, h) =\ & 3.33758 \cdot \sqrt{sf} + 2.442582 \cdot \frac{1}{\sqrt{h}} - 3.072612 \cdot \left( \sqrt{sf} \cdot \frac{1}{\sqrt{h}} \right) \\
& - 2.505019 + \mathit{DMD}(1) \cdot \left( 1.87948 - 1.689478 \cdot \sqrt{sf} - 1.865428 \cdot \frac{1}{\sqrt{h}} + 1.664073 \cdot \left( \sqrt{sf} \cdot \frac{1}{\sqrt{h}} \right) \right)
\end{split}
\end{equation}

\normalsize

\begin{figure}[htbp] 
\centering
\includegraphics[scale=0.13]{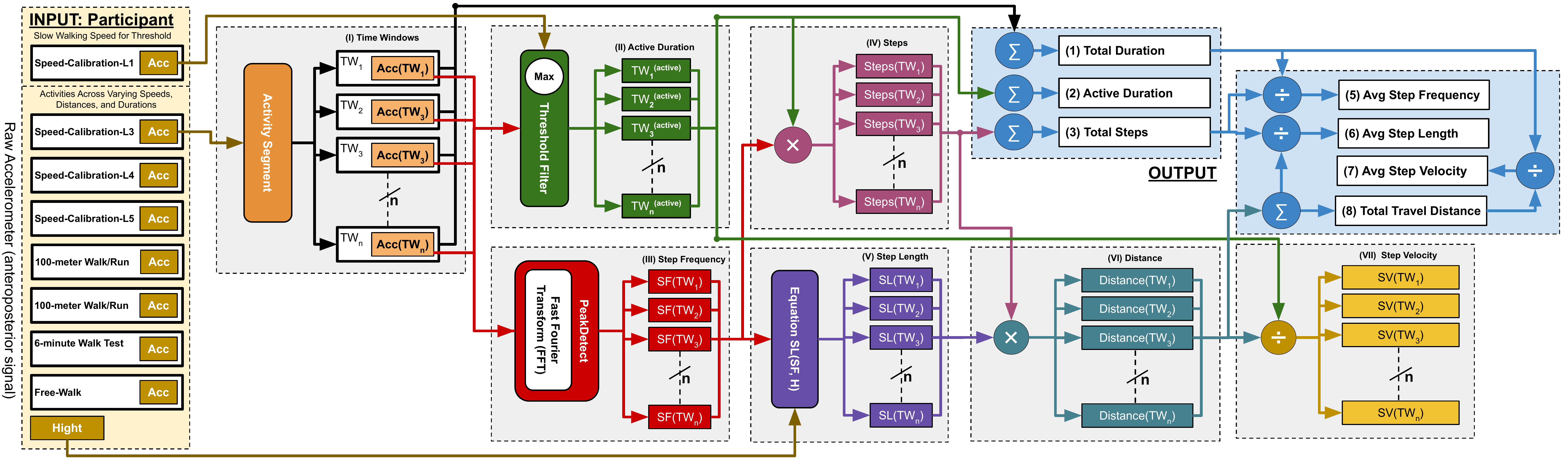}
\caption{Block diagram illustrating the data flow and processing pipeline of our model for gait analysis. The SC-L1 model dynamically adjusts the threshold, while the input consists of acceleration signals from seven unseen gait activities. Key components of the process include step frequency and height, and step length estimation based on Equation~\eqref{EQ3_FORMAT}.}
\label{FIG_DIAGRAM}  
\end{figure}

\section{Experiment-II: Fast Fourier Transform-based estimation of step frequency, step length, step velocity and travel distance from time-windowed raw accelerometer data}
\label{Experiment-II}

\subsection{Materials and~Methods}

\subsubsection{Assessment of participants}
Simultaneously with data collected in Experiment-I (Section-\ref{Experiment-I}) and using the same methods, participants completed longer-distance, mixed speed walking events including a 100-meter walk/run, 6-minute walk test, and a self-selected pace free walk.

\subsubsection{Fourier-Based Detection of Step Frequency and Count}
\label{FTT_SEC}

We developed an iterative process to estimate gait parameters using the anteroposterior (Z-axis) raw accelerometer signal as displayed in Figure~\ref{FIG_DIAGRAM}.  To measure the frequency of steps and estimate the frequency of steps and estimate the count of steps we extracted and estimated the following  values using these steps:
\begin{enumerate}[label=\Roman*.]

\item 
    \label{itm:tw}\textbf{Split the Raw Signal into Time Windows (TW):}  
    We start by collecting the z-axis acceleration data for the activity and then segment the data into 5-second time windows. Within each window, we determine the inactive periods for each individual second in the $TW_{i}$.

    \begin{figure}[htbp] 
    \centering
    \includegraphics[scale=0.39]{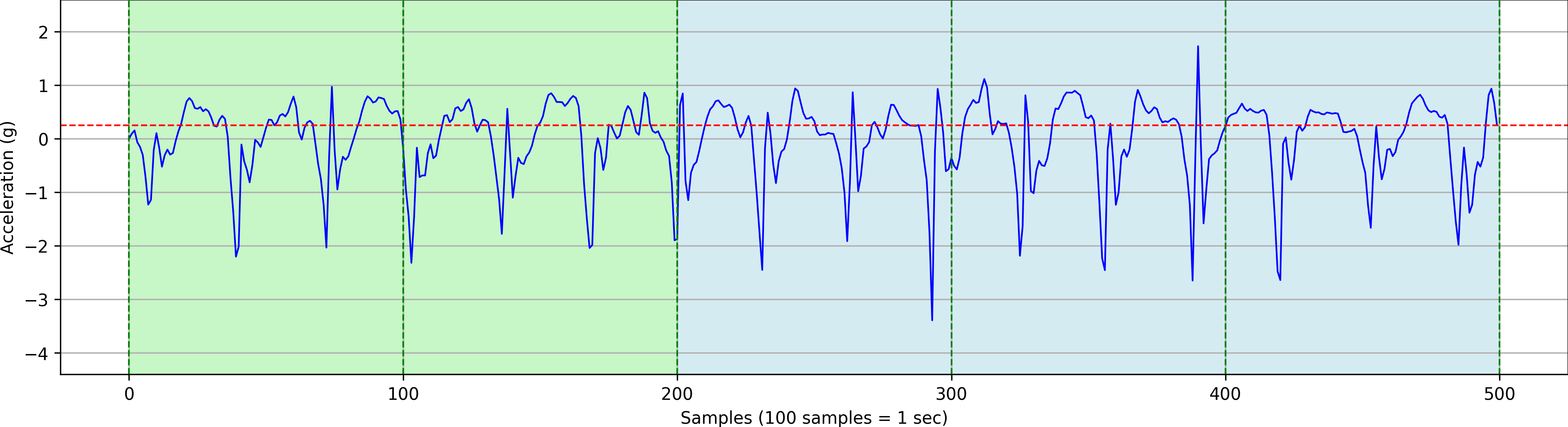}
    \caption{Raw accelerometer data (z-axis) collected during a walking/running trial, demonstrating the time-domain signal collected during ambulation. }
    \label{FIG_RAW}  
    \end{figure}

\item \label{itm:active_time}
    
    \textbf{Determine the Active Duration Time in Each TW}:  
    We collect z-axis acceleration data at the slowest speed (SC-L1 - very slow walking) and compute the threshold, defined as:
    
    \[
    \colorbox{gray!20}{$\text{Threshold} = \mu_{\text{peaks}} + \sigma_{\text{peaks}}$}
    \]
    
    where ${\text{peak}}_i$ denotes the highest peak in a single step at index $i$, with the number of steps in SC-L1 ranging from $i = 1$ to $m$.
    
    \[
    \colorbox{gray!20}{$
    \begin{aligned}
        \mu_{\text{peaks}} &= \frac{1}{m} \sum_{i=1}^{m} \text{peak}_i 
        \quad \text{and} \quad   
        \sigma_{\text{peaks}} &= \sqrt{\frac{1}{m} \sum_{i=1}^{m} (\text{peak}_i - \mu_{\text{peaks}})^2}
    \end{aligned}$}
    \]
    
\begin{figure}[htbp]
    \centering
    \includegraphics[scale=0.28]{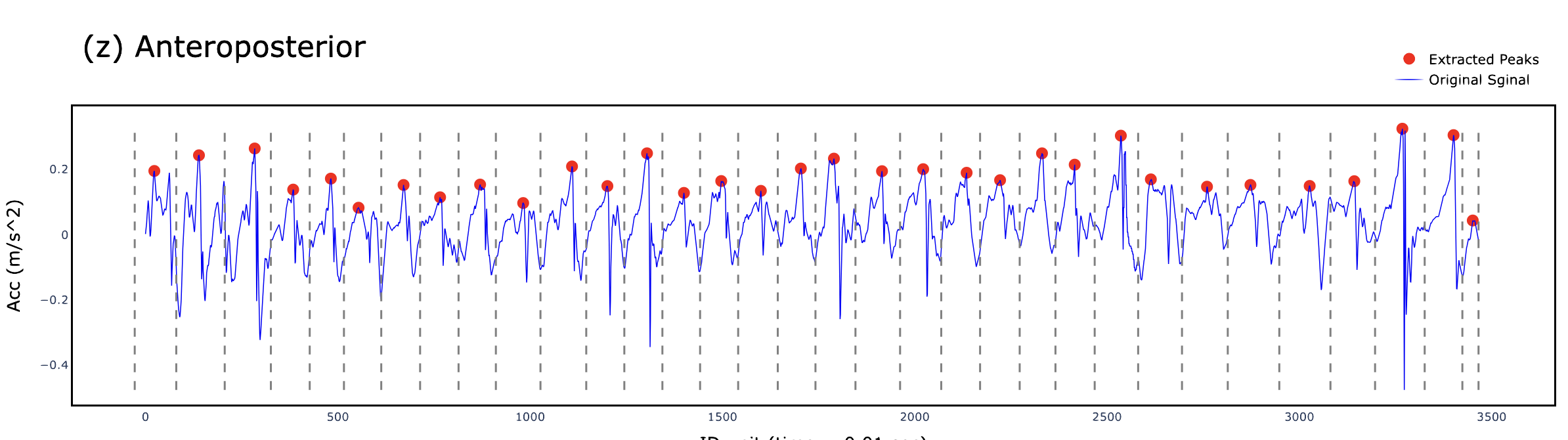}
    \captionsetup{justification=centering} 
    \caption{In SC-L1, the raw acceleration signal is represented by the blue time-series, while the red dots indicate the highest peak in each step. The $\text{peak}_i$ represents the maximum acceleration value in the $i^{th}$ step, where $m$ is the total number of steps in the SC-L1 activity. The gray dotted vertical lines denote the edge duration of each step. These peak values are subsequently used to compute the mean and standard deviation, which serve as the basis for dynamically determining the threshold.}
    \label{FIG_SC-L1}  
\end{figure}

    This threshold, represented by the red horizontal line in Figure~\ref{FIG_ACTIVETIME}, is computed using the mean ($\mu$) and standard deviation ($\sigma$) of peak values in \textbf{SC-L1} as shown in Figure~\ref{FIG_SC-L1}. If the maximum G-force in any 1-second period within the TW is greater than this threshold, that second is classified as \textbf{active time} (highlighted in green). Otherwise, it is classified as \textbf{inactive time} (highlighted in red). This classification rule is applied to every second within the TW.

\[
\colorbox{gray!20}{%
\ensuremath{
\begin{aligned}
    \text{TW}_i^{\text{(active)}} =
    \begin{cases}
        \text{TW}_i, & \text{if } \max[\text{Acc(TW$_i$)}] \geq \text{Threshold} \\
        0, & \text{otherwise}
    \end{cases}
\end{aligned}
}}
\]

\begin{figure}[H]
    \centering
    \includegraphics[scale=0.24]{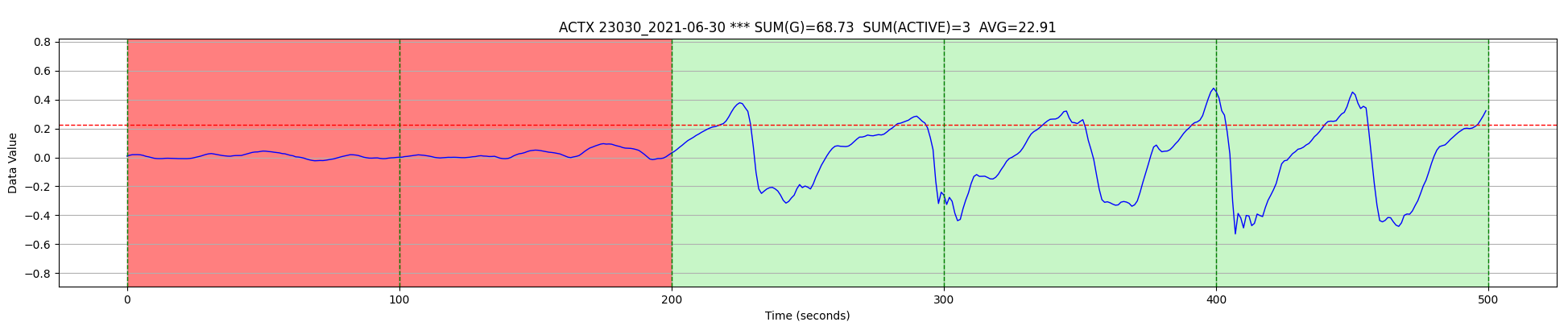}
    \captionsetup{justification=centering} 
    \caption{Determination of active and inactive periods based on peak acceleration thresholds in the slowest walking speed calibration (SC-L1).}
    \label{FIG_ACTIVETIME}  
\end{figure}

\item 
    \textbf{Estimate Step Frequency per TW}:  
For each TW, we convert the z-axis acceleration signal (Figure~\ref{FIG_RAW}) to the frequency domain. Figure~\ref{FIG_FFT_TW} illustrates a frequency spectrum (FFT) outcome for the raw TW. We use the highest peak in the frequency domain to determine the step frequency for each TW, denoted as $\text{Step Frequency}_{\text{TW}}$.

\[
\colorbox{gray!20}{$
\text{Step Frequency}_{\text{TW}_{i}} = \operatorname{PeakDetect} \Big( \operatorname{FFT}[\text{Acc(TW$_{i}$)}] \Big)
$}
\]

\noindent where:
\begin{itemize}
    \item $\text{Acc}_{\text{TW}}$ is the raw acceleration signal recorded within a given TW.
    \item $\operatorname{FFT}(\cdot)$ applies the Fast Fourier Transform (FFT) to convert the time-domain acceleration signal into the frequency domain. For example, Figure~\ref{FIG_FFT_TW} illustrates the frequency domain representation of the TW shown in Figure~\ref{FIG_RAW}.

\begin{figure}[H]
    \centering
    \includegraphics[scale=0.28]{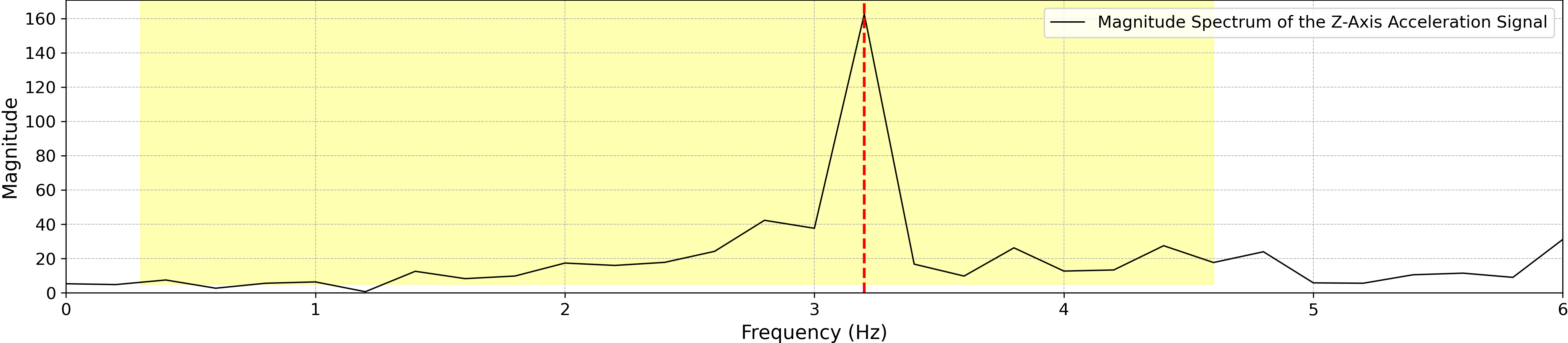}
    \captionsetup{justification=centering} 
    \caption{Fast Fourier Transform (FFT) analysis of accelerometer data within a 5-second time window. The figure highlights the dominant step frequency component extracted for gait cycle estimation.}
    \label{FIG_FFT_TW}  
\end{figure}

    \item $\operatorname{PeakDetect}(\cdot)$ identifies the dominant peak frequency from the transformed signal, corresponding to the step frequency in that TW.
    \item $\text{Step Frequency}_{\text{TW}}$ is the estimated step frequency for the given time window.
\end{itemize}

    In the frequency domain, we identify the highest peak (by magnitude), which represents the step frequency of the selected TW. The following constraints are applied during detection:  
    
    \begin{enumerate}  
    \item Identify the highest peak in the frequency domain within the range of 0.3 to 4.6 Hz; otherwise, the step frequency is set to 0.  
    \item If two peaks exist and the lower-frequency peak is less than $60\%$ of the highest frequency peak, but its magnitude is at least $60\%$ of the highest frequency peak, then the smaller peak is selected as the $\text{Step Frequency}_{\text{TW}}$ as shown in Figure~\ref{FIG_FFT_1ST_PEAK}. Otherwise, the highest frequency peak is determined as the $\text{Step Frequency}_{\text{TW}}$ as shown in Figure~\ref{FIG_FFT_2ND_PEAK}.  

    \begin{figure}[H] 
    \centering
    \includegraphics[scale=0.32]{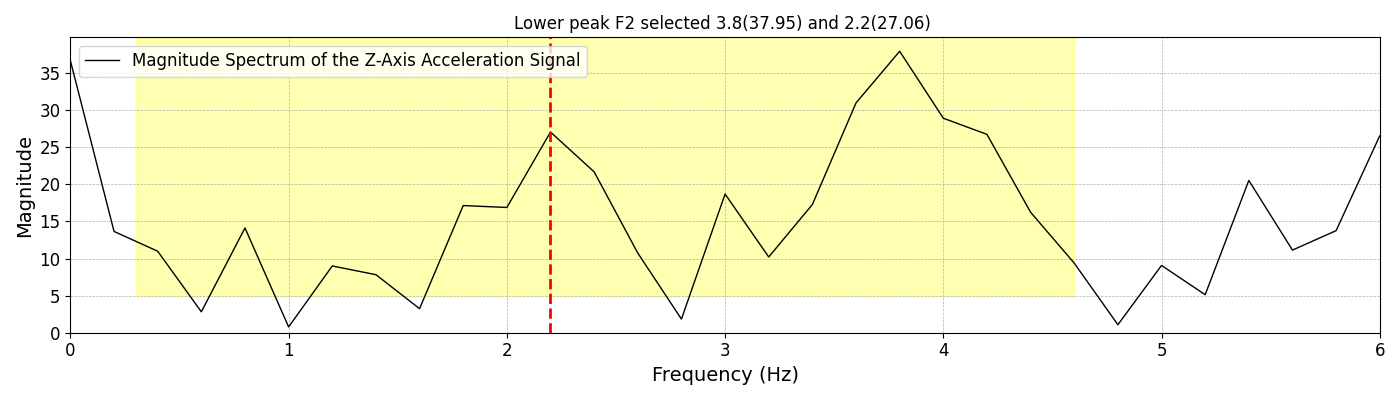}
    \caption{Example of FFT spectrum where the first peak is identified as the step frequency due to its significant magnitude.}
    \label{FIG_FFT_1ST_PEAK}  
    \end{figure}
    
    \begin{figure}[H] 
    \centering
    \includegraphics[scale=0.32]{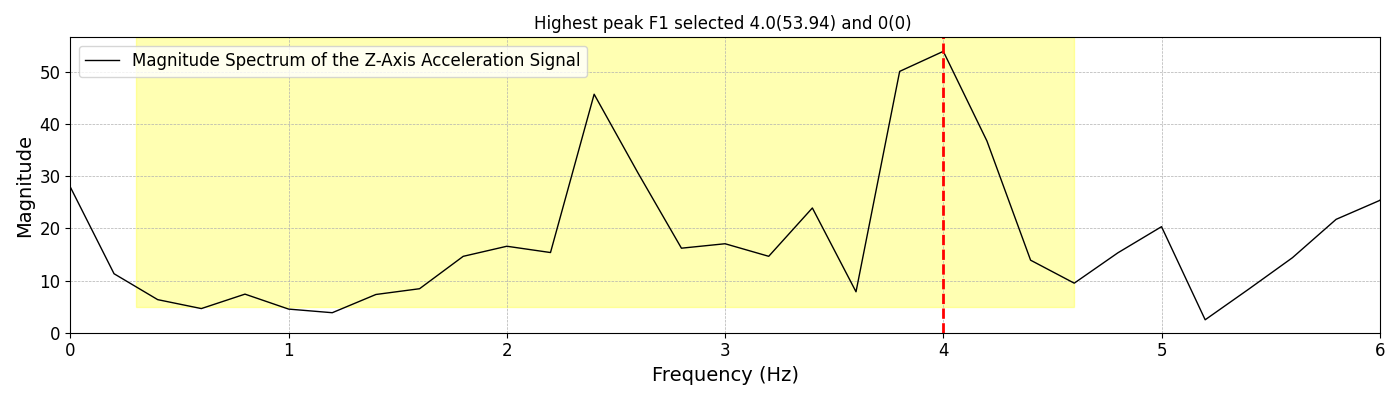}
    \caption{Example of FFT spectrum where the second peak is selected as the step frequency based on predefined criteria.}
    \label{FIG_FFT_2ND_PEAK}  
    \end{figure}

\end{enumerate}

Once the $\text{Step Frequency}_{\text{TW}}$ for a given TW is determined, we calculate the following per TW:

\item \textbf{Estimate the Number of Steps per TW}:  
Multiply the step frequency by the active duration time (as discussed in Section~\ref{FTT_SEC} item \ref{itm:active_time}) within the 5-second time window (0–5 seconds).
\[
\colorbox{gray!20}{$\text{Steps}_{\text{TW}_{i}} = \text{Step Frequency}_{TW_{i}} \times \text{Active Duration}_{TW_{i}}$}
\]

\item \textbf{Estimate Step Length per TW}:  
Estimate the step length for each TW using Equation~\eqref{EQ3_FORMAT}. This calculation is based on the participant’s height and the step frequency derived from the FFT-based step frequency peaks:

\[
\colorbox{gray!20}{$
\begin{aligned}
\text{Step Length}_{\text{TW}_{i}} = SL(SF_{\text{TW}_{i}}, H)
\end{aligned}
$}
\]

where \( SL(sf, h) \) is defined in Equation~\eqref{EQ3_FORMAT}, \( sf_{\text{TW}} \) is the step frequency for the given time window, and \( h \) is the participant's standing height.

\item \textbf{Estimate the Travel Distance per TW}:  
Compute the distance for each TW using:  
\[
\colorbox{gray!20}{$
\begin{aligned}
\text{Distance}_{\text{TW}_{i}} = \text{ Steps}_{TW_{i}} \times \text{Step Length}_{TW_{i}}
\end{aligned}
$}
\]

\item \textbf{Estimate Step Velocity per TW}:  
Calculate the step velocity in meters per second for each TW. The step velocity for each TW is estimated using:  
\[
\colorbox{gray!20}{$
\begin{aligned}
\ensuremath{
\text{Step Velocity}_{\text{TW}_i} = \frac{\text{Distance}_{\text{TW}_i}}{  \text{TW}_i^{\text{(active)}}  } 
}
\end{aligned}
$}
\]

\end{enumerate}

Then, we use the calculated values per TW to estimate the total values:

\begin{enumerate}

\item \textbf{Estimate the Total Duration}:  
Sum the $\text{TW}_i$ across all time windows (TWs) using: 
    \[
    \colorbox{gray!20}{$
    \begin{aligned}
    \text{Total Duration} = \sum_{i=1}^{n} \text{TW}_i
    \end{aligned}
    $}
    \]

    where $n$ represents the total number of time windows (TW), and $\text{TW}_i$ denotes the duration of the $n$-th time window.

\item \textbf{Estimate the Active Duration}:  
Sum the Active Duration $\text{TW}_i^{\text{(active)}}$ across all time windows (TWs) using: 
    \[
    \colorbox{gray!20}{$
    \begin{aligned}
    \text{Active Duration} = \sum_{i=1}^{n} \text{TW}_i^{\text{(active)}}
    \end{aligned}
    $}
    \]

\item \textbf{Estimate the Total Number of Steps}:  
Sum the number of steps across all TWs using:  
\[
\colorbox{gray!20}{$
\begin{aligned}
\text{Total Steps} = \sum_{\text{i}=1}^{n} \text{Steps}_{\text{TW}_{i}}
\end{aligned}
$}
\]
where \(N\) is the total number of time windows, and \(\text{Steps}_{\text{TW}}\) represents the number of steps detected in each individual time window.

\item \textbf{Estimate Average Step Frequency}:  
Compute the average step frequency using:  
\[
\colorbox{gray!20}{$
\begin{aligned}
\text{Avg Step Frequency} = \frac{\text{Total Steps}}{\text{Total Duration}}
\end{aligned}
$}
\]
(Note: The total duration includes both active and inactive time.)

\item \textbf{Estimate Average Step Length}:  
Calculate the average step length using:  
\[
\colorbox{gray!20}{$
\begin{aligned}
\text{Avg Step Length} = \frac{\text{Total Travel Distance}}{\text{Total Steps}}
\end{aligned}
$}
\]

\item \textbf{Estimate Average Step Velocity}:  
Compute the average step velocity using:  
\[
\colorbox{gray!20}{$
\begin{aligned}
\text{Avg Step Velocity} = \frac{\text{Total Travel Distance}}{\text{Total Duration}}
\end{aligned}
$}
\]

\item \textbf{Estimate Total Travel Distance}:  
Sum the travel distances of all TWs to estimate the total travel distance for the entire effort using:  
\[
\colorbox{gray!20}{$
\begin{aligned}
\text{Total Travel Distance} = \sum_{\text{i}=1}^{n} \text{Distance}_{\text{TW}_{i}}
\end{aligned}
$}
\]

\end{enumerate}

\subsubsection{Machine learning-based prediction of gait characteristics using the Walk4Me system}
Accurate and efficient estimation of gait characteristics is essential for mobility assessment and clinical evaluation. The Walk4Me system integrates machine learning (ML) methodologies to predict gait parameters from accelerometer signals obtained via a single waist-worn device. This section provides an overview of our ML-based approach, which has been described in detail in our prior publications~\cite{ramli2023walk4me, ramli2024gait_1,ramli2024gait_2}, and is included here for comparative purposes. While our primary focus in this study is the application of FFT-based analysis due to its suitability for high-volume data processing, the ML approach remains relevant for understanding gait characteristics.

\textbf{Overview of Machine Learning Approach}: Our ML-based method employs supervised learning techniques to estimate key gait parameters, including step count, step length, stride duration, and gait speed. Raw accelerometer signals are preprocessed through feature extraction, followed by classification and regression modeling. The primary workflow includes:

\begin{itemize}
    \item \textbf{Preprocessing and Feature Engineering:} Raw accelerometer data undergoes filtering and transformation to extract temporal and frequency-domain features.
    \item \textbf{Gait Event Detection:} A deep learning-based classifier is utilized to identify gait events, leveraging annotated datasets for model training.
    \item \textbf{Step Length Estimation:} Regression models map peak acceleration values to estimated step lengths using individualized calibration procedures.
    \item \textbf{Distance and Speed Prediction:} By aggregating estimated step lengths and step durations, the system calculates gait speed and total distance traveled.
\end{itemize}

\textbf{Comparative Analysis with FFT-Based Methods}: In prior studies, the ML-based approach demonstrated high accuracy in step detection and gait characterization, particularly in individuals with atypical gait patterns, such as those with Duchenne muscular dystrophy (DMD)~\cite{ramli2024gait_2}. However, FFT-based methods, which focus on spectral decomposition of accelerometer signals, offer advantages in processing efficiency, particularly for large-scale data streams. By converting time-series accelerometer signals into frequency components, FFT enables rapid identification of dominant gait-related frequencies, making it well-suited for automated and high-throughput analysis.

While the Walk4Me system’s ML-based approach has been validated in clinical settings and remains a viable option for gait characterization, FFT methods provide computational efficiency and scalability, particularly for handling high-volume data. The inclusion of ML-based methods in this study serves as a benchmark for comparison, highlighting the trade-offs between data-driven prediction models and frequency-domain analysis.

\subsubsection{Evaluation of precision and bias of model-based estimates compared to ground-truth observation} 
We compared the step count, step frequency, step length, step velocity, and travel distance estimates from the FFA and Walk4Me models to observed measurements. Our validation data included a variety of walking tasks, such as comfortable and fast walks over 10m, 25m, and 100m, a 6-minute walk test, and a free walk. We compared the FFT model and Walk4Me models to the observations, as well as the two models against each other. To assess agreement and bias, we used Bland-Altman analysis to calculate mean percentage differences and limits of agreement. Since walking is cyclical, we normalized residuals and limits of agreement to the percentage of each measurement to account for error increasing proportionally with repetition. We also compared model slopes and intercepts using Passing-Bablok regression, with strong agreement defined as a confidence interval including 1 for slopes and 0 for intercepts~\cite{aaaaa}\cite{Passing1983A}. For acceptable agreement, we chose slopes between between 0.9 and 1.1, and intercepts within 2\% of the maximum observed ground truth value. Additionally, we calculated Lin's Concordance Correlation Coefficient ($\rho_{c}$) to assess both precision and bias, with strong agreement $>$0.95 and a lower limit of 0.9 for acceptable agreement~\cite{lawrence1989concordance}. 

We replicated methods described by Kirk and colleagues~\cite{kirk2024mobilise}, who used mean absolute error (MAE) and mean relative absolute error (MRAE) to describe accuracy of step velocity measurements during structured and daily-living stype activities.  Anticipating that distributions of absolute error would be non-normal, we instead compared median absolute error (MdAE) and median absolute percent error (MdAPE) values and interquartile ranges for absolute and percent error of each model in typical controls, DMD participants, and the combined population for step count, step frequency, step length, step velocity, and travel distance.    

\subsection{Results}

\subsubsection{Participants}
Participant characteristics are identical to Experiment-I (Section-\ref{Experiment-I}), and are shown in Table~\ref{participant_table}. We conducted 325 ambulatory task assessments in 19 participants with DMD and 25 typically-developing controls.  All tasks included ground-truth total distances and elapsed times.  Video recordings required for calculation of step counts, step lengths and step velocities were available for 298 (91\%) of the assessments. 

\subsubsection{Precision and Bias of estimatred gait parameters}
\begin{figure}[htbp] 
\centering
\includegraphics[scale=0.60]{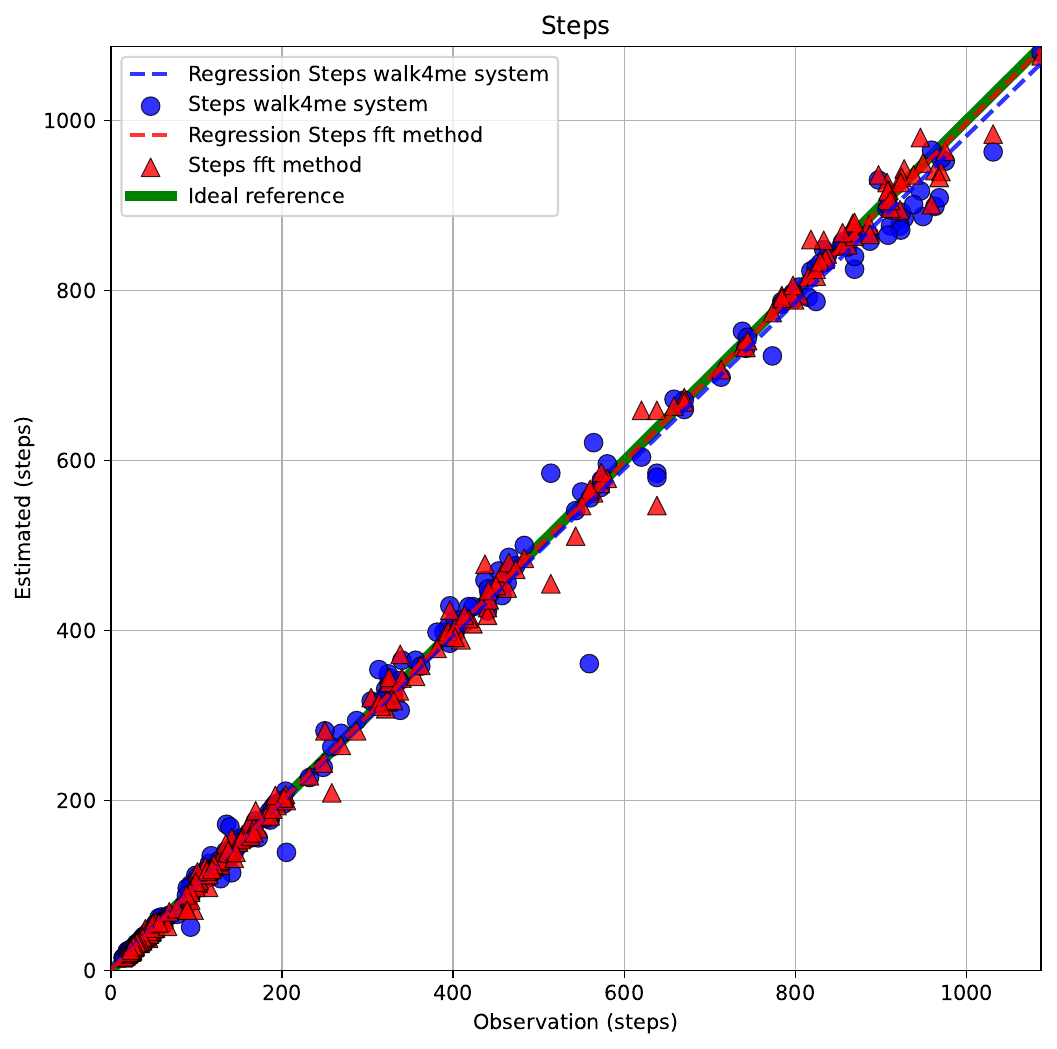}
\caption{Comparison of estimated step counts using FFT-based and Walk4Me models versus ground-truth step counts.}
\label{FIG_ERROR_STEPS}  
\end{figure}

\begin{figure}[htbp]
    \centering
    \begin{subfigure}[b]{0.48\linewidth}
        \centering
        \includegraphics[scale=0.43]{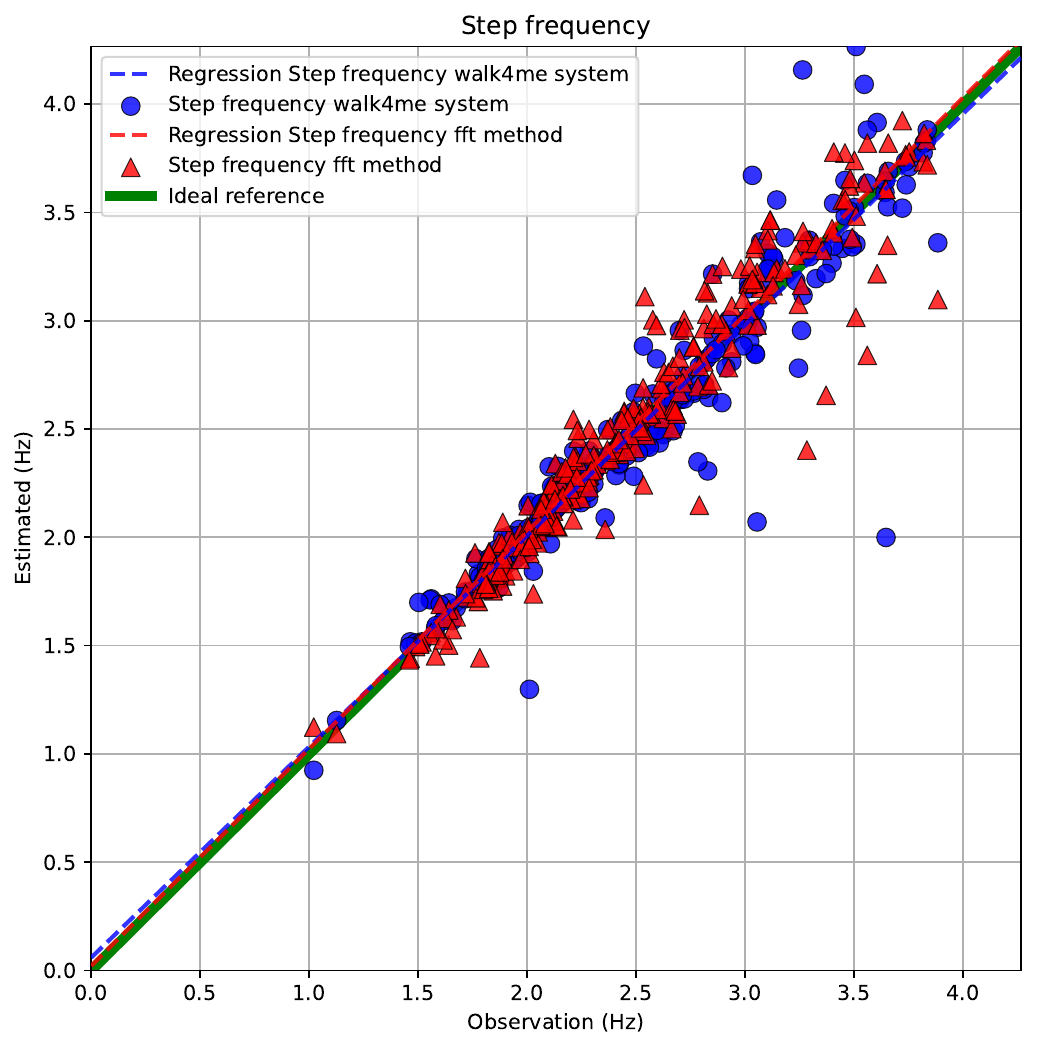}
        \caption{Estimated step frequency.}
        \label{fig:SF}
    \end{subfigure}
    \hfill
    \begin{subfigure}[b]{0.48\linewidth}
        \centering
        \includegraphics[scale=0.43]{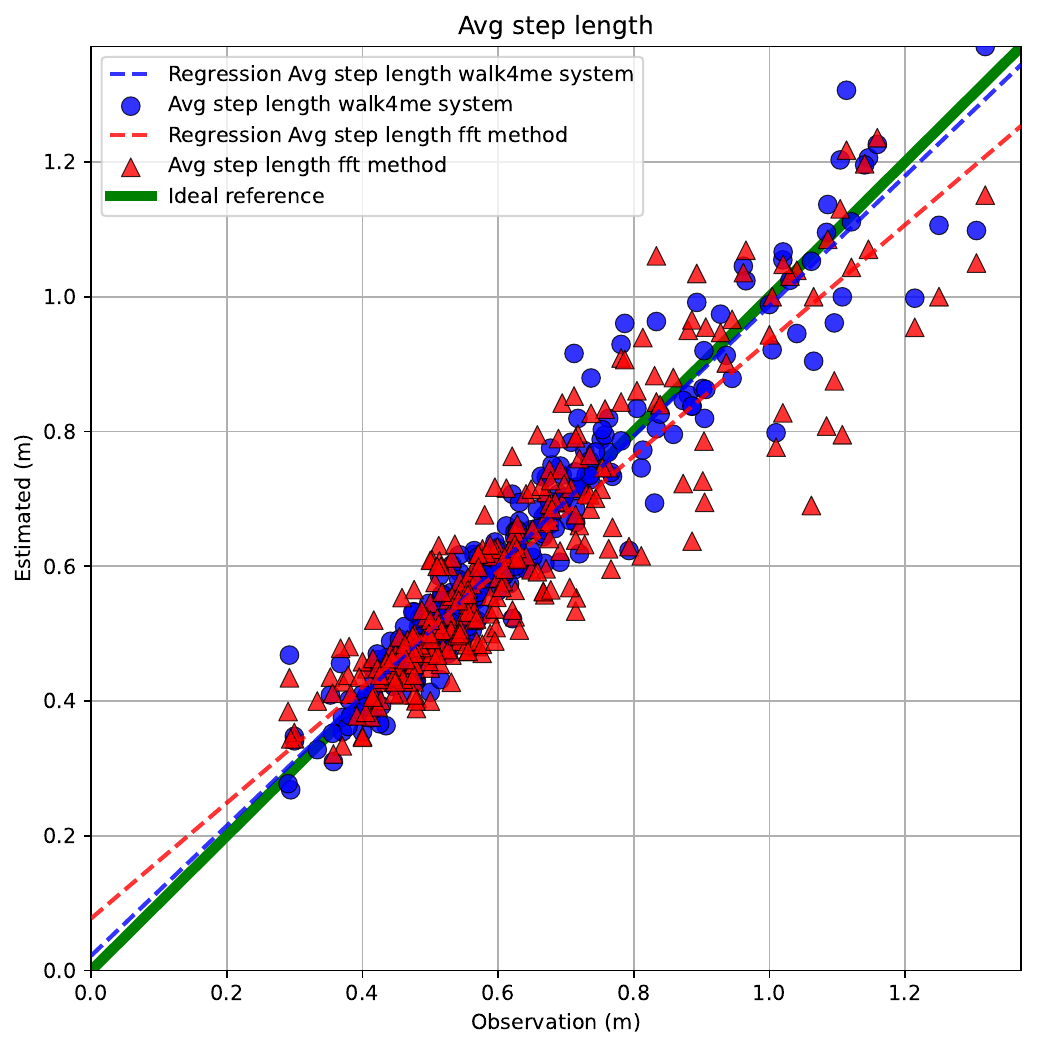}
        \caption{Estimated average step length.}
        \label{fig:SL}
    \end{subfigure}

    \vspace{10pt}

    \begin{subfigure}[b]{0.48\linewidth}
        \centering
        \includegraphics[scale=0.43]{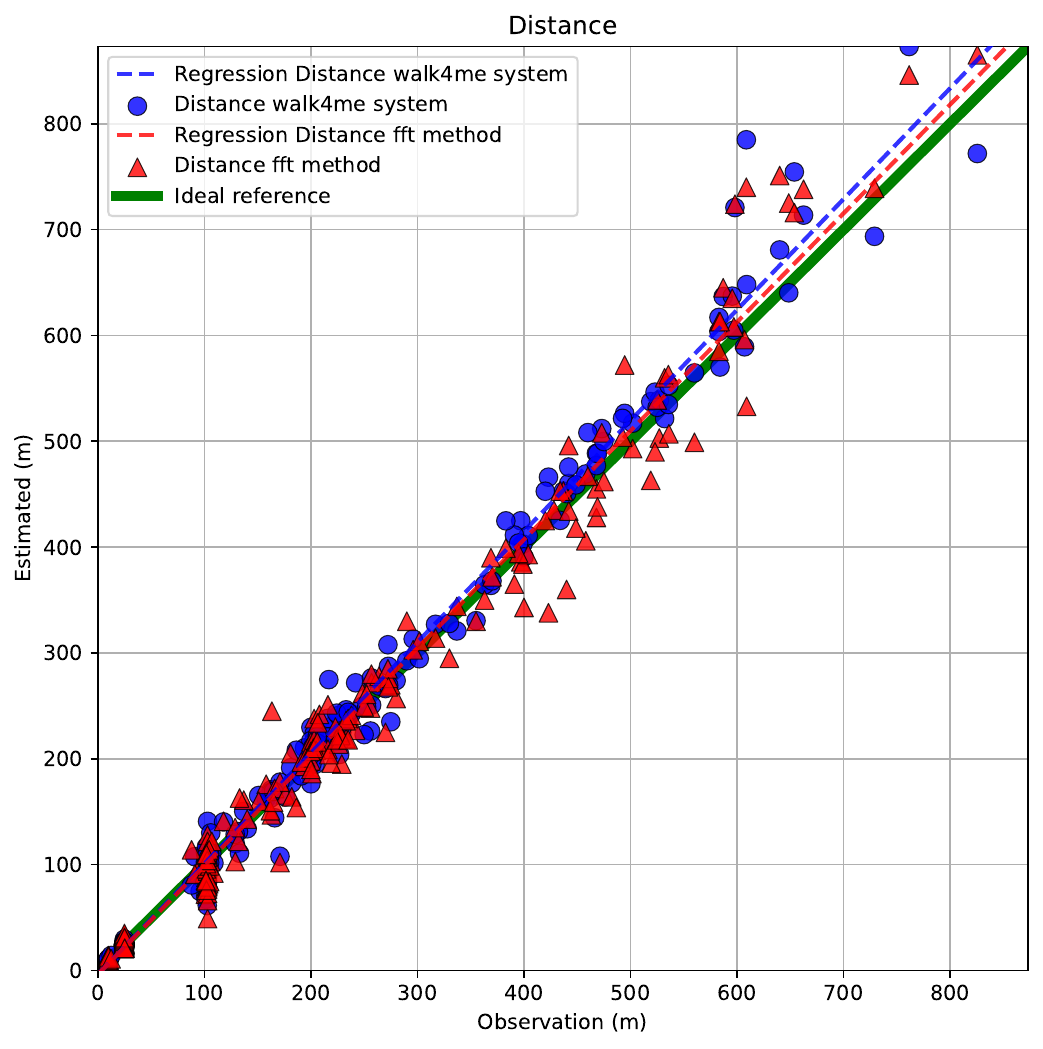}
        \caption{Estimated cumulative distance.}
        \label{fig:DIS}
    \end{subfigure}
    \hfill
    \begin{subfigure}[b]{0.48\linewidth}
        \centering
        \includegraphics[scale=0.43]{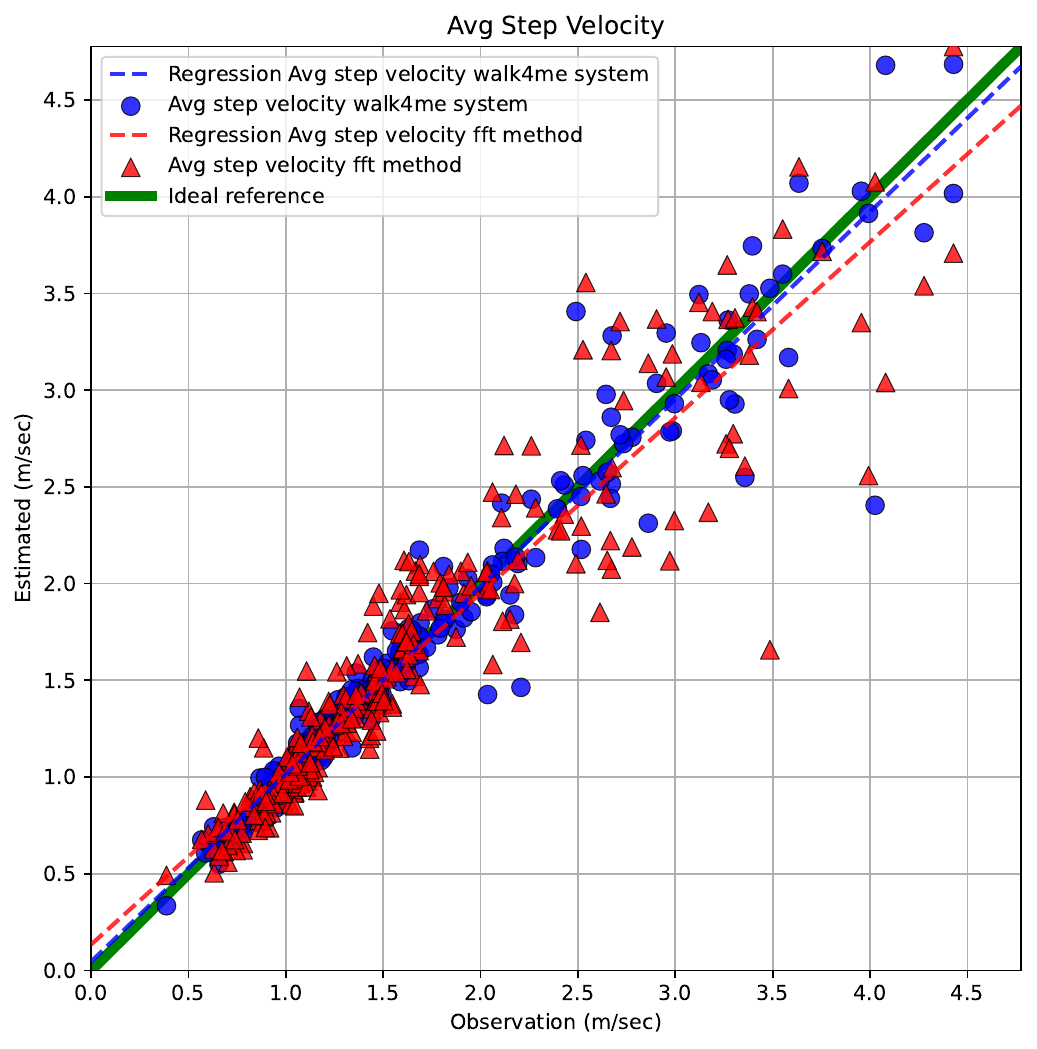}
        \caption{Estimated average step velocity.}
        \label{fig:V}
    \end{subfigure}

    \vspace{5pt}

    \caption{Error analysis for estimated gait parameters across various activities. The subfigures show 
    (a) step frequency,  
    (b) average step length, 
    (c) cumulative travel distance,  
    and (d) average step velocity.}
    \label{FIG_ERROR_ALL}
\end{figure}

FFT-based and Walk4Me-based estimates of step count, step frequency, step length, step velocity and travel distance vs. observed ground-truth measurements are shown in Figure~\ref{FIG_ERROR_STEPS} and Figure~\ref{FIG_ERROR_ALL}. Results of BA, PB and LCC are shown comparing each parameter to observed ground-truth measures in Table~\ref{FFT Method vs. Observed (Ground Truth)}, Table~\ref{Walk4Me System vs. Observed (Ground Truth)}, and  Table~\ref{FFT Method vs. Walk4Me System}.   

\begin{itemize}
    \item \textbf{Step count} comparisons demonstrated excellent agreement with observation and between models, falling within our specified criteria for accuracy, precision and bias.  Bland-Altman regression showed mean percent differences $<$1\% and narrow limits of agreement, while Passing-Bablok regression demonstrated strong agreement in both slopes and intercepts.  Lin’s ($\rho_{c}$) values $>$0.99 indicated high precision and low bias. 
    \item \textbf{Step frequency} comparisons demonstrated excellent agreement with observation and between models, falling within our specified criteria for accuracy, precision and bias.  Mean percent differences were $<$1\% and limits of agreement were narrow, and there was acceptable to strong agreement in both slopes and intercepts.  Lin’s ($\rho_{c}$) values between 0.93 and 0.96 indicated moderate to high precision and correspondingly low bias.  
    \item \textbf{Step length} estimate comparisons demonstrated strong agreement with observation and between models, falling within our specified criteria for accuracy, precision and bias.  Mean percent differences were $<$1.26\% and limits of agreement were slightly wider due to use of group-derived and individually-calibrated models.  There was acceptable to strong agreement in both slopes and intercepts.  Lin’s ($\rho_{c}$) values between 0.91 and 0.96 indicated acceptable to high precision and correspondingly low bias.
    \item \textbf{Distance} comparisons demonstrated excellent agreement with observation and between models, falling within our specified criteria for accuracy, precision and bias.  Mean percent differences were $<$1\% and limits of agreement were slightly wider due to use of group-derived and individually-calibrated models. There was acceptable to strong agreement in both slopes and intercepts.  Lin’s ($\rho_{c}$) values $>$0.99 indicated high precision and low bias. 
    \item \textbf{Step velocity} estimate comparisons demonstrated strong agreement with observation and between models, falling within our specified criteria for accuracy, precision and bias.  Mean percent differences were $<$1\% and limits of agreement were slightly wider due to use of group-derived and individually-calibrated models.  There was acceptable to strong agreement in both slopes and intercepts. Lin’s ($\rho_{c}$) values between 0.93 and 0.98 indicated moderate to high precision and correspondingly low bias. 
\end{itemize}

We evaluated the overall error of the estimates by assessing the Mean Absolute Error (MAE) for stride velocity measurements across all activities. The FFT models showed an MAE (SD) of [0.17 (0.22) m/s], while the Walk4Me models had an MAE of [0.1 (0.18) m/s]. Our results were consistent with those reported by Kirk and colleagues for complex daily activities.  

A comparison of absolute and percentage errors for each metric confirmed that the differences between model-based estimates and ground-truth measurements were not normally distributed (Shapiro-Wilk $<$0.0001). As a result, we present the remaining findings using the median and interquartile ranges for the Median Absolute Error (MdAE) and Median Absolute Percentage Error (MdAPE), as shown in Table~\ref{tab:MdAE} and Table~\ref{tab:MdAPE}. Overall, both estimation techniques performed well, showing acceptable levels of error. 

\begin{table}[H]
\raggedright 
\renewcommand{\arraystretch}{1.2}
\setlength{\tabcolsep}{3pt} 
\caption{\textbf{Precision and accuracy of measurements FFT Method vs. Observed (Ground Truth)}}
\label{FFT Method vs. Observed (Ground Truth)}
\resizebox{1.0\textwidth}{!}{
\begin{tabular}{l c cc c cc c}
\toprule
 \multicolumn{2}{c}{\multirow{2}{*}{\textbf{Metric}}} & \multicolumn{2}{c}{\textbf{PB Regression}} & \multicolumn{2}{c}{\textbf{BA Plot Analysis}} & \multirow{2}{*}{\textbf{Limits of}} & \multirow{2}{*}{\textbf{Lin's Concordance}} \\
\cmidrule(lr){3-4} \cmidrule(lr){5-6} 
 & \textbf{N} & \textbf{Slope (two-sided, 95\% CI)} & \textbf{Intercept (two-sided, 95\% CI)} & \textbf{N} & \textbf{Mean \% Difference (two-sided, 95\% CI)} & \textbf{Agreement (\%)} & \textbf{Correlation} \\
\midrule
Step Count & 297 & 1 (0.99, 1) & 1 (1, 1.06) & 297 & 0.98 (0.32, 1.63) & -10.22, 12.17 & 0.9992 \\
Step Frequency (Hz) & 298 & 1.06 (1.03, 1.08) & -0.1 (-0.16, -0.06) & 298 & 0.92 (0.24, 1.59) & -10.68, 12.52 & 0.9634 \\
Step Length (m) & 298 & 0.99 (0.94, 1.05) & 0 (-0.03, 0.03) & 298 & -1.29 (-2.66, 0.08) & -24.91, 22.32 & 0.9083 \\
Distance (m) & 325 & 1.03 (1.01, 1.04) & -1.13 (-1.4, -0.35) & 325 & -0.64 (-2.11, 0.82) & -27, 25.7 & 0.9917 \\
Step Velocity (m/s) & 297 & 1.05 (1.01, 1.09) & -0.06 (-0.11, -0.01) & 297 & -0.41 (-1.95, 1.13) & -26.79, 25.97 & 0.9395 \\
\bottomrule
\end{tabular}
}
\end{table}

\begin{table}[H]
\raggedright 
\renewcommand{\arraystretch}{1.2}
\setlength{\tabcolsep}{3pt} 
\caption{\textbf{Precision and accuracy of measurements Walk4Me System vs. Observed (Ground Truth)}}
\label{Walk4Me System vs. Observed (Ground Truth)}
\resizebox{1.0\textwidth}{!}{
\begin{tabular}{l c cc c cc c}
\toprule
 \multicolumn{2}{c}{\multirow{2}{*}{\textbf{Metric}}} & \multicolumn{2}{c}{\textbf{PB Regression}} & \multicolumn{2}{c}{\textbf{BA Plot Analysis}} & \multirow{2}{*}{\textbf{Limits of}} & \multirow{2}{*}{\textbf{Lin's Concordance}} \\
\cmidrule(lr){2-4} \cmidrule(lr){5-6} 
 & \textbf{N} & \textbf{Slope (two-sided, 95\% CI)} & \textbf{Intercept (two-sided, 95\% CI)} & \textbf{N} & \textbf{Mean \% Difference (two-sided, 95\% CI)} & \textbf{Agreement (\%)} & \textbf{Correlation} \\
\midrule
Step Count & 297 & 1 (0.99, 1) & 0 (0, 0.2) & 297 & -0.22 (-1, 0.55) & -13.54, 13.09 & 0.9979 \\
Step Frequency (Hz) & 298 & 1 (0.99, 1) & 0 (0, 0.06) & 298 & -0.22 (-1, -0.55) & -13.52, 13.08 & 0.9533 \\
Step Length (m) & 296 & 1.03 (1, 1.06) & -0.02 (-0.03, 0) & 296 & -0.05 (-0.95, 0.86) & -15.55, 15.46 & 0.9633 \\
Distance (m) & 323 & 1.04 (1.03, 1.04) & -1.2 (-1.56, -0.86) & 323 & -0.36 (-1.3, 0.58) & -17.2, 16.48 & 0.9939 \\
Step Velocity (m/s) & 295 & 1.02 (1, 1.04) & -0.02 (-0.05, 0) & 295 & -0.12 (-1.06, 0.81) & -16.11, 15.86 & 0.9754 \\
\bottomrule
\end{tabular}
}
\end{table}

\begin{table}[H]
\raggedright 
\renewcommand{\arraystretch}{1.2}
\setlength{\tabcolsep}{3pt} 
\caption{\textbf{Precision and accuracy of measurements FFT Method vs. Walk4Me System}}
\label{FFT Method vs. Walk4Me System}
\resizebox{1.0\textwidth}{!}{
\begin{tabular}{l c cc c cc c}
\toprule
 \multicolumn{2}{c}{\multirow{2}{*}{\textbf{Metric}}} & \multicolumn{2}{c}{\textbf{PB Regression}} & \multicolumn{2}{c}{\textbf{BA Plot Analysis}} & \multirow{2}{*}{\textbf{Limits of}} & \multirow{2}{*}{\textbf{Lin's Concordance}} \\
\cmidrule(lr){2-4} \cmidrule(lr){5-6} 
 & \textbf{N} & \textbf{Slope (two-sided, 95\% CI)} & \textbf{Intercept (two-sided, 95\% CI)} & \textbf{N} & \textbf{Mean \% Difference (two-sided, 95\% CI)} & \textbf{Agreement (\%)} & \textbf{Correlation} \\
\midrule
Step Count & 325 & 1 (0.99, 1) & -1 (-1, 0.08) & 325 & -0.83 (-1.73, 0.06) & -16.9, 15.24 & 0.9973 \\
Step Frequency (Hz) & 325 & 0.91 (0.89, 0.93) & 0.18 (0.13, 0.22) & 325 & -0.86 (-1.76, 0.03) & -16.9, 15.21 & 0.9343 \\
Step Length (m) & 323 & 1.07 (1.02, 1.11) & 0.03 (-0.06, 0) & 323 & 1.08 (-0.12, 2.28) & -20.38, 22.54 & 0.9316 \\
Distance (m) & 323 & 1 (0.99, 1.03) & -0.24 (-0.79, 0.33) & 323 & 0.34 (-1.18, 1.86) & -26.86, 27.54 & 0.9919 \\
Step Velocity (m/s) & 323 & 0.98 (0.95, 1.02) & 0.02 (-0.03, 0.07) & 323 & 0.34 (-1.18, 1.86) & -26.86, 27.53 & 0.9334 \\
\bottomrule
\end{tabular}
}
\end{table}

The strongest agreement was observed for the directly-measured items step count and step frequency. In these cases, MdAPE estimates were consistently less than 5\% of the ground truth, with the time-windowed FFT-based methods showing a slight advantage in performance. For calculated metrics, FFT-based estimates of step length, distance, and step velocity also showed good agreement, with MdAPE estimates generally under 10\% at walking speeds. Similarly, estimates from the Walk4Me machine learning system outperformed the FFT-based methods, with MdAPE results ranging from 3\% to 10\%. 

In general, error was slightly lower for typical controls compared to patients with Duchenne Muscular Dystrophy (DMD), likely due to the more homogeneous gait in the control group. However, this trend did not hold during running activities, where control participants demonstrated higher error rates for distance and step velocity estimates.

\begin{table}[htbp]
    \centering
    \caption{Median absolute error (MdAE[IQR]) of model-based estimates for activities varying in speed and duration.}
    \resizebox{1.0\textwidth}{!}{
    \begin{tabular}{lcccccc}
        \toprule
        \multirow{2}{*}{\textbf{Comfortable/Fast Walk}} & \multicolumn{2}{c}{Control (n=82)} & \multicolumn{2}{c}{DMD (n=50)} & \multicolumn{2}{c}{All (n=132)} \\
        \cmidrule(lr){2-3} \cmidrule(lr){4-5} \cmidrule(lr){6-7}
        & FFT-based & Walk4Me & FFT-based & Walk4Me & FFT-based & Walk4Me \\
        \midrule
        Step Count (steps) & 1 [1-2] & 1 [0-1] & 1 [1-1] & 0 [0-1] & 1 [1-2] & 0 [0-1] \\
        Step Frequency (Hz) & 0.08 [0.03-0.14] & 0.05 [0-0.11] & 0.11 [0.07-0.15] & 0 [0-0.07] & 0.09 [0.04-0.14] & 0 [0-0.1] \\
        Step Length (m) & 0.04 [0.02-0.08] & 0.02 [0.01-0.05] & 0.05 [0.02-0.08] & 0.02 [0.01-0.03] & 0.04 [0.02-0.08] & 0.02 [0.01-0.04] \\
        Distance (m) & 2 [1-4] & 0.69 [0.34-1.18] & 1 [0.8-2] & 0.3 [0.14-0.51] & 1 [0.9-3] & 0.49 [0.2-1.03] \\
        Step Velocity (m/s) & 0.11 [0.04-0.32] & 0.04 [0.02-0.07] & 0.10 [0.06-0.15] & 0.03 [0.01-0.05] & 0.11 [0.04-0.23] & 0.04 [0.02-0.07] \\
        \midrule
        \multirow{2}{*}{\textbf{Free Walk}} & \multicolumn{2}{c}{Control (n=34)} & \multicolumn{2}{c}{DMD (n=17)} & \multicolumn{2}{c}{All (n=51)} \\
        \cmidrule(lr){2-3} \cmidrule(lr){4-5} \cmidrule(lr){6-7}
        & FFT-based & Walk4Me & FFT-based & Walk4Me & FFT-based & Walk4Me \\
        \midrule
        Step Count (steps) & 10 [4-15] & 7.5 [4-16] & 5 [3-9.5] & 13 [8-21] & 6 [4-15] & 10 [4-17] \\
        Step Frequency (Hz) & 0.05 [0.02-0.08] & 0.03 [0.02-0.08] & 0.02 [0.02-0.05] & 0.05 [0.04-0.08] & 0.03 [0.02-0.08] & 0.04 [0.02-0.08] \\
        Step Length (m) & 0.02 [0.01-0.04] & 0.02 [0.01-0.04] & 0.04 [0.03-0.09] & 0.03 [0.01-0.05] & 0.03 [0.01-0.05] & 0.02 [0.01-0.04] \\
        Distance (m) & 10 [7-15] & 9.27 [4.7-14.9] & 16 [6-24] & 10.5 [5.6-21.8] & 10.5 [6.1-23] & 9.4 [4.9-17.2] \\
        Step Velocity (m/s) & 0.05 [0.03-0.08] & 0.04 [0.02-0.08] & 0.08 [0.03-0.12] & 0.06 [0.03-0.11] & 0.06 [0.03-0.09] & 0.05 [0.02-0.08] \\
        \midrule
        \multirow{2}{*}{\textbf{6-minute Walk Test}} & \multicolumn{2}{c}{Control (n=35)} & \multicolumn{2}{c}{DMD (n=19)} & \multicolumn{2}{c}{All (n=54)} \\
        \cmidrule(lr){2-3} \cmidrule(lr){4-5} \cmidrule(lr){6-7}
        & FFT-based & Walk4Me & FFT-based & Walk4Me & FFT-based & Walk4Me \\
        \midrule
        Step Count (steps) & 9 [5-22] & 16 [5-42] & 7 [2-11] & 10 [1-32] & 9 [4-20] & 15 [3-37] \\
        Step Frequency (Hz) & 0.02 [0.01-0.06] & 0.06 [0.01-0.12] & 0.02 [0.01-0.03] & 0.03 [0-0.09] & 0.02 [0.01-0.06] & 0.04 [0.01-0.1] \\
        Step Length (m) & 0.04 [0.02-0.08] & 0.04 [0.02-0.06] & 0.02 [0.01-0.04] & 0.02 [0.01-0.04] & 0.03 [0.01-0.06] & 0.03 [0.01-0.06] \\
        Distance (m) & 30 [13-61] & 23.4 [9.9-41.6] & 15 [7-26] & 13.8 [6.5-23.7] & 24.5 [11-53] & 18.4 [8.8-35.4] \\
        Step Velocity (m/s) & 0.08 [0.03-0.17] & 0.07 [0.03-0.11] & 0.04 [0.01-0.07] & 0.04 [0.02-0.08] & 0.07 [0.03-0.12] & 0.06 [0.02-0.11] 
        \\
        \midrule
        \multirow{2}{*}{\textbf{100M walk/run}} & \multicolumn{2}{c}{Control (n=41)} & \multicolumn{2}{c}{DMD (n=20)} & \multicolumn{2}{c}{All (n=61)} \\
        \cmidrule(lr){2-3} \cmidrule(lr){4-5} \cmidrule(lr){6-7}
        & FFT-based & Walk4Me & FFT-based & Walk4Me & FFT-based & Walk4Me \\
        \midrule
        Step Count (steps) & 4 [1-10] & 5 [3-9] & 3 [2-5] & 4 [2-8] & 4 [2-7] & 5 [2-9] \\
        Step Frequency (Hz) & 0.11 [0.04-0.3] & 0.13 [0.07-0.31] & 0.04 [0.03-0.09] & 0.05 [0.04-0.11] & 0.09 [0.03-0.18] & 0.11 [0.05-0.23] \\
        Step Length (m) & 0.07 [0.03-0.17] & 0.05 [0.02-0.07] & 0.05 [0.01-0.09] & 0.03 [0.02-0.05] & 0.07 [0.02-0.14] & 0.04 [0.02-0.06] \\
        Distance (m) & 11 [6-23] & 4.9 [2.8-11.9] & 9 [3-16] & 3.7 [1.9-6.7] & 10 [3.3-20] & 4.2 [2.2-10] \\
        Step Velocity (m/s) & 0.34 [0.14-0.59] & 0.12 [0.08-0.37] & 0.17 [0.04-0.23] & 0.06 [0.03-0.13] & 0.22 [0.09-0.51] & 0.1 [0.05-0.34] \\
        \midrule
        \multirow{2}{*}{\textbf{All Tasks}} & \multicolumn{2}{c}{Control (n=192)} & \multicolumn{2}{c}{DMD (n=105)} & \multicolumn{2}{c}{All (n=297)} \\
        \cmidrule(lr){2-3} \cmidrule(lr){4-5} \cmidrule(lr){6-7}
        & FFT-based & Walk4Me & FFT-based & Walk4Me & FFT-based & Walk4Me \\
        \midrule
        Step Count (steps) & 3 [1-9] & 3 [1-10] & 2 [1-5] & 1 [0-9] & 2 [1-7] & 2 [0-10] \\
        Step Frequency (Hz) & 0.07 [0.02-0.13] & 0.06 [0.01-0.13] & 0.07 [0.02-0.13] & 0.03 [0-0.09] & 0.07 [0.02-0.13] & 0.05 [0-0.11] \\
        Step Length (m) & 0.04 [0.02-0.08] & 0.03 [0.01-0.05] & 0.04 [0.02-0.08] & 0.02 [0.01-0.04] & 0.04 [0.02-0.08] & 0.03 [0.01-0.05] \\
        Distance (m) & 7 [2-16.5] & 2.24 [0.8-13] & 3 [1-16] & 2.4 [0.35-10] & 6 [1.5-16] & 3.1 [0.59-11.94] \\
        Step Velocity (m/s) & 0.09 [0.04-0.27] & 0.06 [0.02-0.11] & 0.08 [0.03-0.16] & 0.04 [0.02-0.08] & 0.09 [0.04-0.21] & 0.05 [0.02-0.09] \\
        \bottomrule
    \end{tabular}}
    \label{tab:MdAE}
\end{table}

\begin{table}[htbp]
    \centering
    \caption{Median absolute percent error (MdAPE[IQR]) of model-based estimates for activities varying in speed and duration.}
    \resizebox{1.0\textwidth}{!}{
    \begin{tabular}{lcccccc}
        \toprule
        \multirow{2}{*}{\textbf{Comfortable/Fast Walk}} & \multicolumn{2}{c}{Control (n=82)} & \multicolumn{2}{c}{DMD (n=50)} & \multicolumn{2}{c}{All (n=132)} \\
        \cmidrule(lr){2-3} \cmidrule(lr){4-5} \cmidrule(lr){6-7}
        & FFT-based & Walk4Me & FFT-based & Walk4Me & FFT-based & Walk4Me \\
        \midrule
        Step Count (\%) & 3.5 [1.5-5.4] & 2 [0-4] & 4.9 [2-6.7] & 0 [0-3.3] & 4.3 [1.6-5.6] & 0 [0-3.8] \\
        Step Frequency (\%) & 3.4 [1.5-5.4] & 2 [0-4] & 5.1 [3.7-6.6] & 0 [0-3.3] & 4.3 [1.8-5.6] & 0 [0-3.8] \\
        Step Length (\%) & 7.8 [3.5-10.7] & 3.8 [1.3-6.9] & 10 [4.5-15] & 3.1 [1.6-7.1] & 8.3 [4-14.7] & 3.5 [1.4-6.9] \\
        Distance (\%) & 8 [4-16] & 2.9 [1.4-4.8] & 10 [8.1-18.4] & 2.8 [1.5-4.9] & 10 [4-16] & 2.9 [1.4-4.8] \\
        Step Velocity (\%) & 8 [4-16] & 2.9 [1.4-4.8] & 10 [8.1-18.4] & 2.8 [1.5-4.9] & 10 [4-16] & 2.9 [1.4-4.9] \\
        \midrule
        \multirow{2}{*}{\textbf{Free Walk}} & \multicolumn{2}{c}{Control (n=34)} & \multicolumn{2}{c}{DMD (n=16)} & \multicolumn{2}{c}{All (n=50)} \\
        \cmidrule(lr){2-3} \cmidrule(lr){4-5} \cmidrule(lr){6-7}
        & FFT-based & Walk4Me & FFT-based & Walk4Me & FFT-based & Walk4Me \\
        \midrule
        Step Count (\%) & 2.4 [1.1-4.3] & 1.8 [0.7-3.9] & 1.2 [0.7-2.4] & 3.0 [2.4-6.1] & 1.6 [0.9-4.0] & 2.5 [1.2-4.0] \\
        Step Frequency (\%) & 2.4 [1.1-4.3] & 1.8 [0.7-3.9] & 1.3 [0.8-2.5] & 2.5 [2.3-4.5] & 1.7 [0.1-4.3] & 2.4 [1.1-4.0] \\
        Step Length (\%) & 5.3 [2.1-7.2] & 3.7 [1.7-6.6] & 14.5 [5.9-17.9] & 5.7 [4.3-10] & 6.0 [2.3-10] & 4.3 [1.7-8.3] \\
        Distance (\%) & 4.7 [3.1-7.3] & 4.7 [2.2-6.8] & 8 [3.1-17.4] & 7 [3.4-13] & 5.4 [3.1-10.6] & 5 [2.7-9.4] \\
        Step Velocity (\%) & 4.7 [3.1-7.3] & 4.2 [1.7-6.8] & 9.7 [3.4-18.5] & 7.7 [3.6-12.6] & 5.2 [3.1-10.5] & 4.8 [2.4-8.4] \\
        \midrule
        \multirow{2}{*}{\textbf{6-minute Walk Test}} & \multicolumn{2}{c}{Control (n=35)} & \multicolumn{2}{c}{DMD (n=19)} & \multicolumn{2}{c}{All (n=54)} \\
        \cmidrule(lr){2-3} \cmidrule(lr){4-5} \cmidrule(lr){6-7}
        & FFT-based & Walk4Me & FFT-based & Walk4Me & FFT-based & Walk4Me \\
        \midrule
        Step Count (\%) & 1.1 [0.6-2.4] & 2.3 [0.6-4.5] & 0.9 [0.4-1.3] & 1.5 [0.1-3.5] & 1.0 [0.5-2.2] & 2.0 [0.4-2.5] \\
        Step Frequency (\%) & 1.1 [0.1-2.4] & 2.3 [0.6-4.5] & 0.9 [0.4-1.4] & 1.5 [0.1-3.5] & 1.0 [0.6-2.3] & 2.0 [0.4-4.5] \\
        Step Length (\%) & 5.6 [3.1-11] & 6 [2.9-10.5] & 3.9 [1.9-7.8] & 3.9 [1.8-8.8] & 5.5 [2.4-10.1] & 5.5 [2-10] \\
        Distance (\%) & 5.4 [2.7-11.1] & 4.5 [2-7.6] & 3.7 [1.8-7] & 3.4 [2.2-6.9] & 5.2 [2.2-10] & 4.1 [2.2-7] \\
        Step Velocity (\%) & 5.2 [2.3-10.9] & 4.9 [2-7.6] & 3.6 [1.2-6.6] & 4.2 [1.7-7.8] & 5.1 [1.8-9.5] & 4.6 [2-7.6] \\
        \midrule
        \multirow{2}{*}{\textbf{100M walk/run}} & \multicolumn{2}{c}{Control (n=41)} & \multicolumn{2}{c}{DMD (n=20)} & \multicolumn{2}{c}{All (n=61)} \\
        \cmidrule(lr){2-3} \cmidrule(lr){4-5} \cmidrule(lr){6-7}
        & FFT-based & Walk4Me & FFT-based & Walk4Me & FFT-based & Walk4Me \\
        \midrule
        Step Count (\%) & 3.1 [1.1-8.3] & 4.3 [2.1-8.9] & 1.5 [1.2-2.8] & 2.3 [1.1-3.9] & 2.9 [1.1-5.2] & 3.5 [1.3-8.6] \\
        Step Frequency (\%) & 3.1 [1.1-8.3] & 4.3 [2.1-8.9] & 1.5 [1.2-2.8] & 2.3 [1.1-3.9] & 2.9 [1.1-5.2] & 3.5 [1.3-8.6] \\
        Step Length (\%) & 8.3 [2.8-20.1] & 4.7 [2.9-7.9] & 8.2 [1.9-17] & 5.5 [3.2-9.1] & 8.2 [2.5-19] & 4.8 [2.9-8.2] \\
        Distance (\%) & 10.7 [5.7-22] & 4.7 [2.6-11.6] & 8.7 [2.9-15.5] & 3.5 [1.9-6.6] & 9.7 [3.2-19.4] & 4.1 [2.2-9.7] \\
        Step Velocity (\%) & 10.7 [5.7-22.3] & 4.7 [2.7-11.6] & 8.7 [2.9-15.5] & 3.7 [1.9-9.1] & 9.7 [3.4-19.4] & 4.2 [2.6-10.3] \\
        \midrule
        \multirow{2}{*}{\textbf{All Tasks}} & \multicolumn{2}{c}{Control (n=192)} & \multicolumn{2}{c}{DMD (n=105)} & \multicolumn{2}{c}{All (n=297)} \\
        \cmidrule(lr){2-3} \cmidrule(lr){4-5} \cmidrule(lr){6-7}
        & FFT-based & Walk4Me & FFT-based & Walk4Me & FFT-based & Walk4Me \\
        \midrule
        Step Count (\%) & 2.7 [0.9-4.9] & 2.4 [0.4-4.5] & 2.2 [0.6-5.6] & 1.3 [0-3.7] & 2.5 [0.9-5.1] & 2.1 [0-4.4] \\
        Step Frequency (\%) & 2.7 [0.9-4.9] & 2.4 [0.4-4.5] & 2.5 [0.9-5.5] & 1.2 [0-3.7] & 2.6 [1-5.2] & 2.1 [0-4.4] \\
        Step Length (\%) & 6.3 [2.8-11.6] & 4.3 [1.7-7.9] & 9.7 [4-15.6] & 4.1 [1.7-8.7] & 7.1 [3.6-14.7] & 4.2 [1.7-8.1] \\
        Distance (\%) & 6.7 [3.2-15] & 3.9 [1.9-6.8] & 10 [2.8-14.8] & 3.5 [1.8-7.3] & 8 [3.1-14.8] & 3.7 [1.9-7] \\
        Step Velocity (\%) & 6.8 [3.2-14.8] & 3.8 [1.8-6.7] & 10 [2.8-15.5] & 3.5 [1.8-7.5] & 8 [3.1-15.5] & 3.7 [1.8-7] \\
        \bottomrule
    \end{tabular}}
    \label{tab:MdAPE}
\end{table}

\section{Discussion}
Here we present a model-based approach to accurately estimate step length in children aged 3 to 16 years with Duchenne muscular dystrophy and typical controls using only step frequency and standing height.  When combined with step counts and elapsed total time, the formula presented here is suitable for creating estimates of travel distance for that time interval.  The equation, which we created from observations of step frequency and stride length across a typical range of slow to fast walking speeds in toddlers, children and teens with Duchenne muscular dystrophy and similarly-aged typically developing controls, may also be suitable for use as a predictive equation for DMD and typical populations, and to create contrasts between the two at different stages of growth and development.   

We further demonstrate that the Fast Fourier Transformation is a computationally efficient and suitable method to estimate step frequency from time-windowed raw signals from inertial measurement units in commercially-available smart phones and other wearable devices during longer-duration recording of daily activities.  Daily step activity patterns and step/stride velocity of people with DMD have been well described~\cite{bjornson2011walking,bjornson2014walking,bjornson2014relation,mcdonald2005utility,mcdonald2005use,servais2022stride,poleur2021normative,rabbia2024stride,servais2024evidentiary}, but less is known about daily distance traveled, and adding estimates of travel distance and velocity may provide new insights into temporal patterns of habitual activity.  The total daily step count decreases only gradually as people with DMD progress in their disease, reducing the utility of this metric as a clinical trial outcome measure.  However, the drop in high-frequency and high-velocity steps becomes more apparent with loss of ability.  Measurement of the 95th percentile of habitual stride velocity – the speed below which 95 percent of steps occur – in the community using ankle-worn IMUs is well-developed and is approved as an outcome measure for clinical trials~\cite{servais2021first, servais2024evidentiary, lilien2019home, poleur2021normative, servais2022stride, rabbia2024stride}.  Yet, velocity of steps or strides alone tells an incomplete story that accounts for neither the length of those steps, nor distances traveled during daily activities, nor the overall patterns of temporospatial gait characteristics over time.   

The absolute error of walking velocity measurements with single IMU-based wearable devices has been reported to increase with the complexity of community walking activities~\cite{kirk2024mobilise}.  In a recent report by Kirk and colleagues, they described absolute error in walking velocity across multiple patient groups of 0.11 m/s (range 0.09 to 0.13 m/s) and relative absolute error of 20.3\% (range 15.48 to 24.88\%) for a combined set of simple and complex gait tasks reflective of real-world walking bouts.  Using similar methods, we report similar if slightly improved levels of accuracy, adding a controlled running task in toddlers, children, and teens that demonstrates an acceptable level of accuracy using single, waist-worn IMU devices.

\begin{figure}[htbp] 
\centering
\includegraphics[scale=0.65, trim=0 280 0 0, clip]{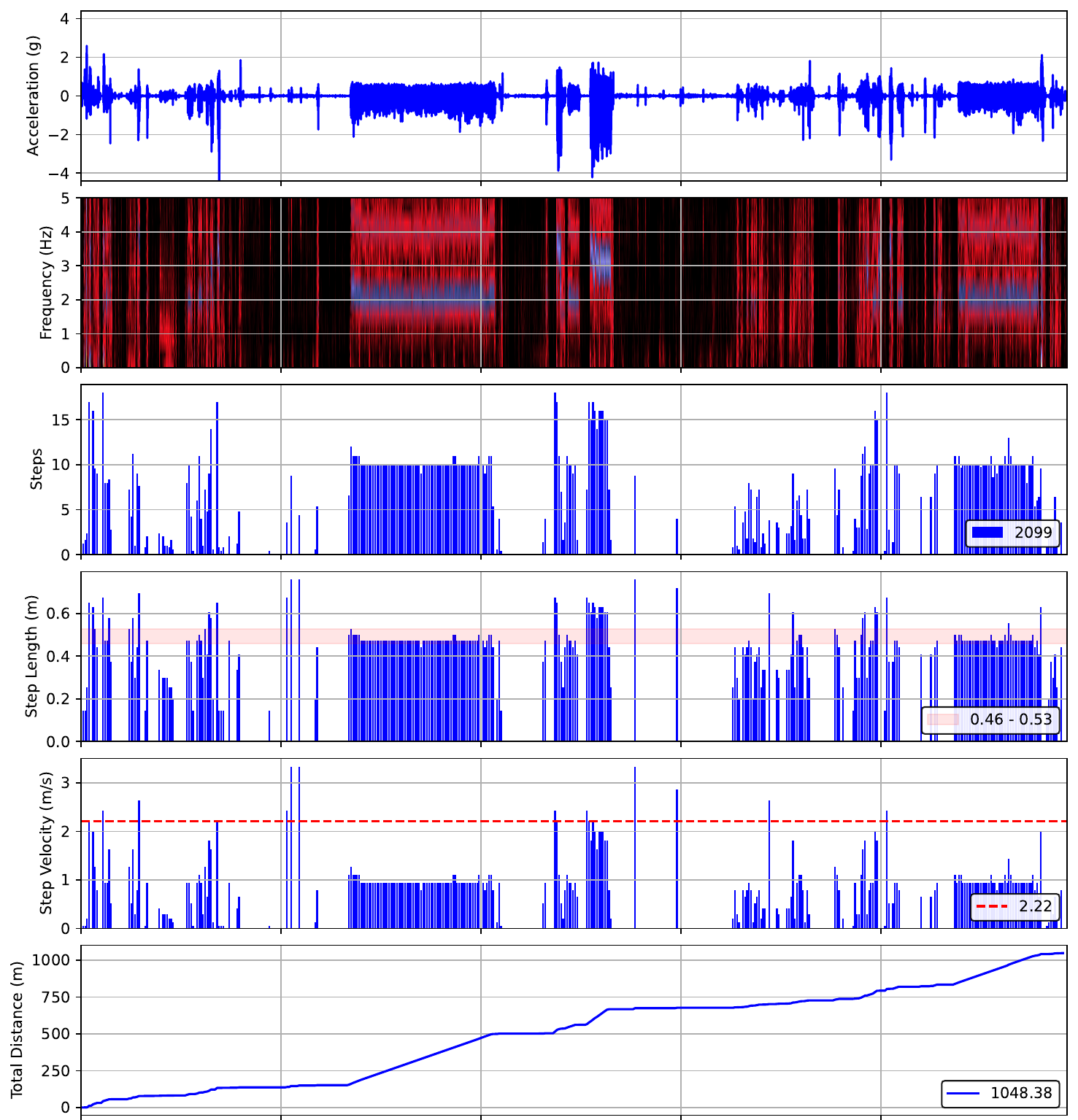}
\caption{Visualization of time-series gait analysis for an extended activity session, showing step frequency, estimated number of steps, estimated step length, step velocity, and cumulative travel distance.}
\label{FIG_RESULT}  
\end{figure}

Used alone or in combination, our two approaches to IMU data analysis allow us to accurately estimate clinically-meaningful gait parameters using a variety of consumer-level, single-sensor devices.  The overall agreement between our FFT-based and individually-calibrated, AI-based Walk4Me models~\cite{ramli2024gait_1,ramli2024gait_2,ramli2023walk4me} indicates that FFT-based estimates can provide an accurate first-pass analysis to identify key events of interest.  As illustrated in Figure 12 depicting data extracted from a 50-minute mixed-task activity using FFT-based analysis and Equation 1, the ability to rapidly generate temporal maps of long-term activity can help us to identify events of interest such as prolonged periods of continuous moderate activity or shorter bouts of high-level burst activity. Here we display the raw accelerometer signal over time, followed by a time-series FFT visual map of step frequency in 5-second time windows.  For each time window, use of Equation 1 then determines step length, shown with a band indicating 35-40 percent of standing height that is common during self-selected pace to fast walking.  Those time-windowed step length estimates yield average velocities for each step, with the dotted red line indicating the 95th percentile step velocity for the recorded activity. Simultaneously, step count and step length yield estimates of cumulative distance traveled, which ultimately provides the basis for evaluation of community-based travel over extended periods of time.  Notably, the areas of highest step frequency are easily visible highlighted in purple in the time-series FFT panel.  These areas represent the 6-minute walk test (between approximately 700s and 1100s) and 10-meter, 25-meter and 100-meter walk/run tests (approximately 1200s to 1400s) and the "free walk" (approximately 2200s to 2500s).  In each case, the last panel illustrates both cumulative distance represented on the Y axis and also velocity represented as the slope of the line.

Once such events are identified, we can then employ more computationally-intensive AI tools to extract an extended panel of clinically-relevant gait features to reflect not only stride length and velocity characteristics, but proportions of forces indicating forward, lateral and vertical travel that are known to be significantly altered in individuals with a variety of pathologic gait patters.  Likewise, extension of measurement tools into the frequency domain using continuous sensor data and tools like FFT opens up the entire spectral analysis toolbox, enabling us to construct time-series maps of activity that quantitatively and visually combine both frequency and power of gait events, and to extract features from such data for inclusion in future predictive classical machine learning and deep learning AI models.

\section{Conclusions}

This study introduces a novel method for estimating step length, step velocity, and travel distance in children with Duchenne Muscular Dystrophy (DMD) and typically developing controls using frequency-domain analysis and regression-based modeling. By leveraging Fast Fourier Transform (FFT)-derived step frequency from single, waist-worn consumer-grade accelerometers, our approach enables accurate estimation of gait parameters in real-world settings.

Our results demonstrate that step frequency, when combined with standing height, provides a reliable predictor of step length in both DMD and typically developing children. The proposed model achieves high accuracy (\textit{R$^2$} = 0.92) and low error (RMSE = 0.06), validating its applicability in community gait evaluation. Additionally, comparisons between FFT-based methods and AI-driven Walk4Me models indicate strong agreement across multiple gait metrics, reinforcing the reliability of FFT as a computationally efficient alternative for large-scale data processing.

Beyond laboratory settings, this method enhances the feasibility of community-based gait monitoring using widely available consumer devices. The ability to quantify step length, velocity, and travel distance without the need for expensive, custom-built IMUs opens new possibilities for monitoring disease progression in DMD, assessing intervention efficacy, and supporting clinical decision-making in mobility-impaired populations.

Future work will focus on refining our models through larger datasets, expanding analysis to additional movement disorders, and integrating real-time data processing capabilities into mobile health applications. By improving accessibility and scalability, our approach has the potential to contribute significantly to gait assessment and digital health monitoring in both clinical and everyday environments.

\section{Supplementary Materials}  
A portion of the source code and a demonstration associated with this paper, along with additional results, are available as of March 31, 2025, at \url{https://albara.ramli.net/research/ff}.

\vspace{6pt}

\section*{Author Contributions}
\begin{itemize}
    \item \textbf{E.K.H.:} Conceptualization, Methodology, Software, Formal analysis, Writing, Supervision, Funding acquisition, Investigation.
    \item \textbf{A.A.R.:} Conceptualization, Methodology, Software, Formal analysis, Writing, Supervision, Validation, Visualization, Investigation, Data Curation.
\end{itemize}
All authors have read and agreed to the published version of the manuscript.

\section*{Funding}
This research was funded by the US Department of Defense (grant number W81XWH-17-1-0477), the Muscular Dystrophy Association (grant number 646805), and intramural pilot funds from the University of California Center for Information Technology Research in the Interest of Society (CITRIS) and the Banatao Institute.

\section*{Institutional Review Board Statement}
The study was conducted according to the guidelines of the Declaration of Helsinki and approved by the Institutional Review Board of the University of California Davis (IRB\#1305174, March 18, 2019).

\section*{Informed Consent Statement}
Written informed consent was obtained from each participant prior to the initiation of study procedures.

\section*{Data Availability Statement}
Due to the human subject and health information privacy nature of the data and our institutional regulations, we will share information upon presentation of evidence of IRB or ethics board review and completion of appropriate data transfer agreements. Requests for data access can be addressed to the corresponding author.





\section*{Acknowledgments}
We would like to thank the participants who contributed their time and data to our study. We also thank the faculty, staff, and students of the UC Davis PM\&R Neuromuscular Research Lab, including Dr. Craig McDonald, Adrell Argonza, Kelly Berndt, Smriti Davey, Erica Goude, Lynea Kaethler, Amanda Lopez, Roch Monachan, Alina Nicorici, David Rodriguez, and Jane Wang, who assisted with participant enrollment, data collection, and data management.

\section*{Conflicts of Interest}
The authors declare no conflict of interest.








\bibliographystyle{unsrt}
\bibliography{cas-refs}

\end{document}